\documentclass[conference,a4paper]{IEEEtran}
\IEEEoverridecommandlockouts
\usepackage{cite}
\usepackage{amsmath,amssymb,amsfonts}
\usepackage{algorithmic}
\usepackage{graphicx}
\usepackage{textcomp}
\usepackage{xcolor}
\usepackage{hyperref}
\usepackage{booktabs}
\usepackage{siunitx}

\def\BibTeX{{\rm B\kern-.05em{\sc i\kern-.025em b}\kern-.08em
    T\kern-.1667em\lower.7ex\hbox{E}\kern-.125emX}}
\begin{document}

\title{Learning To Defer To A Population With Limited Demonstrations}

\author{
Nilesh Ramgolam$^{*}$,
Gustavo Carneiro$^{\dagger}$,
Hsiang-Ting Chen$^{*}$ \\
\\
$^{\dagger}$Centre for Vision, Speech and Signal Processing, University of Surrey, UK \\
$^{*}$Australian Institute for Machine Learning (AIML), University of Adelaide, Australia \\
\\
$^{\dagger}$\texttt{g.carneiro@surrey.ac.uk} \\
$^{*}$\texttt{\{nilesh.ramgolam,tim.chen\}@adelaide.edu.au}
}

\maketitle

\begin{abstract}
This paper addresses the critical data scarcity that hinders the practical deployment of learning to defer (L2D) systems to the population. We introduce a context-aware, semi-supervised framework that uses meta-learning to generate expert-specific embeddings from only a few demonstrations. 
We demonstrate the efficacy of a dual-purpose mechanism, where these embeddings are used first to generate a large corpus of pseudo-labels for training, and subsequently to enable on-the-fly adaptation to new experts at test-time. 
The experiment results on three different datasets confirm that a model trained on these synthetic labels rapidly approaches oracle-level performance, validating the data efficiency of our approach. 
By resolving a key training bottleneck, this work makes adaptive L2D systems more practical and scalable, paving the way for human-AI collaboration in real-world environments. \footnote{To facilitate reproducibility and address implementation details not covered in the main text, we provide our source code and training configurations \href{https://github.com/nil123532/learning-to-defer-to-a-population-with-limited-demonstrations}{here}.}
\end{abstract}


\section{Introduction}
Recent advances in AI systems have achieved near-human or superhuman performance across diverse fields, such as computer vision \cite{farrell2021} and medical image analysis \cite{rajpurkar2017,topol2019,petashvili2024learning,pitawela2025cloc}. Despite these achievements, purely automated AI models often fall short in safety-critical areas, including healthcare diagnostics \cite{Cremer2021,raghu_2019}. This limitation has motivated the emergence of \textit{hybrid intelligence systems} that integrate human experts with AI, leveraging the complementary strengths of both \cite{kamar_2016,dellermann2019,akata2020}. A prominent branch of hybrid intelligence is \textit{Learning to Defer (L2D)}, which enables AI models to either autonomously predict or defer uncertain and high-risk decisions to human experts \cite{madras2018predict,husseinmozannar2020consistent,nguyen2025probabilistic,zhang2024learning,10.5555/3666122.3666281,verma-L2DMultiple,verma2022calibrated}. 

Conventional L2D systems, trained on a fixed cohort of experts, exhibit poor generalization to new individuals at test-time \cite{raghu_2019,verma-L2DMultiple}. To address this, adaptive L2D approaches have emerged that learn to model diverse expert behaviors by conditioning on past decisions effectively using them as a form of context. \cite{tailor2024learning,strong2025expert}. However, the efficacy of these adaptive models is contingent on extensive labeled datasets that capture the full spectrum of population behavior—a requirement posing a significant practical barrier.
Existing methods to mitigate this data dependency are themselves insufficient. They are either architecturally confined to single-expert scenarios \cite{hemmer2023} or, if population-based, cannot generalize to experts unseen during training \cite{nguyen2025probabilistic}. A critical gap therefore exists: there is no data-efficient methodology for training L2D models that are simultaneously context-aware and adaptive to new experts.

To bridge this gap, we propose a context-aware semi-supervised L2D framework that adapts to unseen experts from limited demonstrations. We formulate this as a meta-learning task where the model learns to generate an expert-specific embedding from just a few examples of their decisions, thereby capturing an individual's unique behavior. 
This context-aware embedding serves two critical functions. 
First, during training, it is leveraged to generate a large corpus of pseudo-labels for a diverse population. This synthetically-labeled data then provides the supervision required to train a robust downstream L2D model. 
Second, at test-time, the embedding itself acts as the context vector, enabling the trained L2D model to adapt its deferral strategy to any new expert on the fly.

To summarize, our contributions include:
\begin{itemize}
    \item The introduction of a novel context-aware SSL framework that generates pseudo-labels representing the diverse labeling behaviors within a population of experts, from a handful of labeled examples per expert.
    \item A meta-learning framework that models the meta-expert representation enabling the downstream adaptive L2D models.
    \item Empirical demonstration of our framework's effectiveness across multiple tasks highlighting its ability to achieve robust deferral performance by adapting to new experts even with extremely limited data.
\end{itemize}
Collectively, our work represents a significant advancement in hybrid intelligence, making L2D systems more practical, scalable, and adaptive to real-world collaborative environments.

\begin{figure*}[t]
  \centering
  \includegraphics[width=\textwidth]{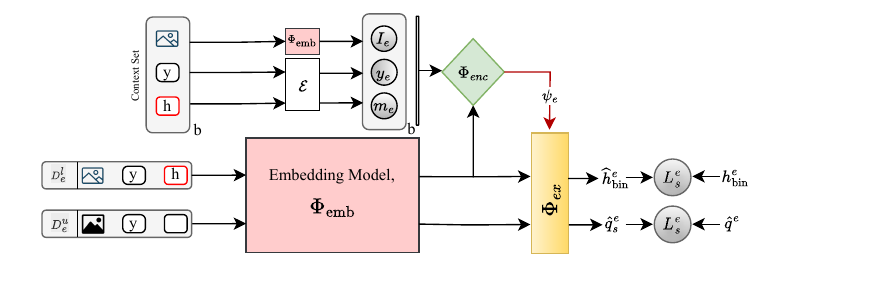}
  \caption{%
    A small context set
    $\mathcal{C}_{e}= \{(x_b,y_b,h_b^{e})\}_{b=1}^{B}$
    summarises expert~$e$.
    Image features from the frozen backbone
    $\Phi_{\text{emb}}$
    are concatenated with label embeddings and processed by the set encoder
    $\Phi_{\text{enc}}$,
    yielding the expert embedding $\boldsymbol{\psi}_{e}$.
    For a query image $x_j$ the same $\Phi_{\text{emb}}$ encodes the image;
    this vector is combined with $\boldsymbol{\psi}_{e}$ and passed to the
    expert-aware head $\Phi_{\text{ex}}$, which predicts whether the expert will
    label the query \emph{correctly} or \emph{incorrectly}.
    Each expert contributes a supervised loss
    $\mathcal{L}_{s}^{e}$ on its labelled data and a consistency loss
    $\mathcal{L}_{u}^{e}$ on its unlabelled data;averaging across these terms yields the
    meta-objective~$\mathcal{L}$ that trains the entire multi-task model across all
    experts.}
  \label{fig:schema}
\end{figure*}

\section{Related Work}
Research over the last few years has investigated combining human and AI capabilities for superior team performance, exceeding what either could achieve alone. Within this context, L2D algorithms, which allow an AI model to either make a prediction or defer to a human expert, have shown promise. Early approaches focused on estimating confidence levels for both the classifier and the expert, deferring when the human was deemed more confident \cite{raghu_2019}. Others optimized for overall team performance by training the classifier to specifically complement the expert's abilities \cite{wilder2020learning}, often jointly training the classifier and deferral mechanism \cite{madras2018predict, okati_2021_differentiable, wilder2020learning}. Further advancements include objective functions with theoretical guarantees for regression \cite{de2020regression}, as well as Mozannar and Sontag’s (2020) consistent surrogate loss inspired by cost-sensitive learning. Building on this, Raman and Yee (2021) personalised the deferral system to specific experts through fine-tuning \cite{raman2021improving}. Additional studies have explored instance assignment under bandit feedback \cite{gao2021human} and allocation within teams comprising an AI and multiple experts \cite{hemmer2022forming}. However, a common assumption underlying much of this work is that the human expert providing input remains the same between the training and deployment phases.

Although L2D-Pop \cite{tailor2024learning} successfully adapts to evolving expert behaviors at test time, its meta-training phase still depends on a vast corpus of labeled data—both ground-truth annotations and expert demonstrations. Acquiring such extensive expert labels is often impractical and prohibitively expensive. One might consider model-based imputation \cite{hemmer2023} to synthesize labels for each expert, but these methods are inherently tailored to a single, fixed expert and would demand costly training of separate models for each individual. This gap highlights the absence of any scalable approach capable of generating pseudo-labels that reflect the full diversity of an expert population. To overcome this limitation, we introduce a meta-learning framework that, given only a small context set from an expert, produces high-quality, expert-aligned pseudo-labels on the fly—using a single, unified model—thereby dramatically reducing the annotation burden for downstream population-aware L2D applications.

\section{Problem Formulation}

To formalize our approach, we define our learning environment as follows. Let the input space be denoted by \(X\) and the set of \(k\) discrete classes by \(Y=\{1,\dots,k\}\). We assume access to a primary ground-truth dataset \(D_{\text{gt}}=\{(x_i,y_i)\}_{i=1}^{N}\). We consider a population of \(M\) experts, \(E=\{e_1,\dots,e_M\}\), where for each expert \(e \in E\), we only have access to a \emph{limited set of historical annotations}, \(D_e^l=\{(x_i,y_ih_i^e)\}_{i\in L}\), where \(h_i^e \in Y\) is the label provided by expert \(e\). The core constraint is that for every expert, this labeled set is small: \(|L| \ll N\).

Training an independent predictive model for each of the \(M\) experts, as in single-expert strategy explored in ~\cite{hemmer2023}, is impractical for a population. This approach would require maintaining \(M\) separate models, prevent sharing statistical strength across related experts, and be operationally expensive. Instead, our goal is to develop a \emph{single, unified model} that can generalize across the entire population \(E\).

To enable this, in our approach our model is conditioned on an expert-specific \emph{context set}, \(C_e\), which provides a few examples of the expert's annotation behaviour. This set is formed by sampling \(B\) instances from the expert's available data, \(D^l_e\), where \(B\) is a small, fixed-size hyperparameter. For each sampled instance, we form a triplet containing the input, its ground truth, and the expert's label: \(C_e=\{(x_b,y_b,h_b^e)\}_{b=1}^{B}\). Each triplet \((x_b,y_b,h_b^e)\) allows the model to observe the conditional error pattern of the expert—that is, the discrepancy between their label \(h_b^e\) and the ground-truth \(y_b\) for a given input \(x_b\).

 \section{Approach}
\label{sec:methodology}
To address per-expert data scarcity, we propose a semi-supervised framework that learns to predict expert behavior by conditioning on a small set of their past decisions. Our approach uses a model-based meta-learning paradigm \cite{mishra2017simple} to generate an expert-specific embedding that captures an individual's unique behavioral style. This embedding serves a dual purpose: it is used to generate a large corpus of pseudo-labels to train a downstream L2D model, and it enables on-the-fly adaptation to unseen experts at test-time.

As depicted in Figure~\ref{fig:schema}, our architecture generates expert‐aligned pseudo‐labels via three modules. The Context Set Encoder ($\Phi_{\mathrm{enc}}$) first creates a behavioral embedding from an expert’s context set. This expert embedding, together with input features from the Embedding Model ($\Phi_{\mathrm{emb}}$) is then passed to the Expert Predictor ($\Phi_{\mathrm{ex}}$) which produces the final, tailored pseudo‐label. The encoder ($\Phi_{\mathrm{enc}}$) is also re-used at test time to adapt the L2D model to new experts. We explain the process in more detail in the following sections.

\subsection{Foundational Feature Representation}
\label{sec:feature-repr}
The first step in our pipeline is to establish a shared feature representation for all images. We begin by establishing a foundational feature representation. Let \(\Phi_{\text{emb}}\colon X \rightarrow \mathbb{R}^{f}\) be a feature extractor that embeds high-dimensional image inputs into a compact representation. This corresponds to the Embedding Model, \(\Phi_{\text{emb}}\) block in Figure \ref{fig:schema}. Following standard transfer-learning protocols, we pre-train \(\Phi_{\text{emb}}\) jointly with a classification head \(\Omega\colon \mathbb{R}^{f}\rightarrow Y\) on the complete ground-truth dataset
\[
D_{\text{gt}} = \{(x_i, y_i)\}_{i=1}^N
\]
using the cross-entropy loss \(\mathcal{H}\):
\[
\mathcal{L}_{\text{emb}}
= \frac{1}{N}
\sum_{i=1}^{N}
\mathcal{H}\bigl(
y_i,\,
\Omega\bigl(\Phi_{\text{emb}}(x_i)\bigr)
\bigr).
\]
This pre-training phase yields a robust feature space that underpins the subsequent, more nuanced task of modelling individual expert behaviour. The parameters of \(\Phi_{\text{emb}}\) are thereafter frozen.
\subsection{Context-Aware Expert Predictor}
\label{sec:context-model}
With this foundational feature space established, we now introduce the core of our method, which learns to model expert-specific behaviours. The core of our method comprises two jointly trained components: an expert context set encoder \(\Phi_{\text{enc}}\) and an expert predictor model \(\Phi_{\text{ex}}\). This system uses an attention-based architecture to condition its predictions on a concise summary of an expert’s past decisions.

\medskip\noindent
\textbf{Expert context set encoder.} The first component, \(\Phi_{\text{enc}}\), is responsible for processing the expert's history. As shown at the top of Figure~\ref{fig:schema}, it takes the expert’s context set \(\mathcal{C}_e\), where each element consists of an image, its ground-truth label, and the expert’s label. For every example in the set, the encoder forms an initial representation by concatenating the image features (from \(\Phi_{\text{emb}}\)) with embeddings of the two labels. These representations are then fed through a self-attention mechanism, enabling the model to capture the internal relationships within the expert’s historical decisions. 

\medskip\noindent
\textbf{Expert predictor model.} The second component is an expert-aware behaviour predictor, \(\Phi_{\text{ex}}\), which makes the final binary correctness prediction of an expert. This is achieved via a cross-attention operation where the query image’s feature vector attends to the context-aware vectors produced by \(\Phi_{\text{enc}}\). This step, illustrated in the center of Figure~\ref{fig:schema}, produces the single, query-specific expert embedding \({\psi}_e\), which emphasizes the most relevant past decisions for the current query.

Finally, the expert predictor \(\Phi_{\text{ex}}\) uses this embedding $\psi_e$ to predict the correctness of the expert's annotation for the new query instance \(x^*\). This is done by concatenating the embedding \(\psi_e\) and the query image’s features before passing it to the prediction head \(\Phi_{\text{ex}}\), which yields final the binary outcome:
\[
\hat{h}^e_{bin}
= \Phi_{\text{ex}}\bigl(x^*,\,\psi_e\bigr)
\;\in\;\{0,1\},
\]
where \(1\) signifies that the expert will predict a correct label for instance $x^*$.

\subsection{Semi-Supervised Learning}\label{sec:training}
We train a single model to predict behaviour for all \(E\) experts in a multi-task  gsetting, jointly minimising supervised and unsupervised objectives.\footnote{Weak and strong augmentations are denoted \(\mathrm{Aug}_w(\cdot)\) and \(\mathrm{Aug}_s(\cdot)\), respectively, following FixMatch~\cite{sohn2020fixmatch}.}

\paragraph{Data splits.}
The ground-truth dataset \(D_{\text{gt}}\) is partitioned by annotation availability.  
The small \emph{annotated} subset is
\[
D^{l}=\Bigl\{(x_i,y_i)\,\bigm|\,
       \exists\,e\in E\ \text{s.t.}\ (x_i,h_i^{e})\in D^{l}_e\Bigr\},
\]
while the \emph{unannotated} subset is \(D^{u}=D_{\text{gt}}\setminus D^{l}\).  
With \(\ell=\lvert D^{l}\rvert\) and \(u=\lvert D^{u}\rvert\), we have \(N=\ell+u\) total training instances.

\paragraph{Supervised loss.}
For expert \(e\), define the binary target \(h_{\text{bin},i}^{e}=1\) if \(h_{i}^{e}=y_i\) and \(0\) otherwise.  
The supervised learning objective for expert $e$ is
\begin{equation}
\mathcal{L}_{s}^{e}
  =\frac{1}{\ell}
    \sum_{i=1}^{\ell}
      \mathcal{H}\!\Bigl(
        h_{\text{bin},i}^{e},\,
        \Phi_{\text{ex}}\!\bigl(
          \Phi_{\text{emb}}\!\bigl(\mathrm{Aug}_{w}(x_i)\bigr),\,
          \boldsymbol{\psi}_{e}
        \bigr)
      \Bigr),
\label{eq:Ls}
\end{equation}
where \(\boldsymbol{\psi}_{e}\) is the expert embedding.

\paragraph{Unsupervised consistency loss}
For the larger unannotated set \(D_u\), we generate pseudo-labels and enforce a consistency objective to improve the model’s generalization. The core idea is that the model should predict the same outcome for an image even when that image is subjected to strong distortion. First, we create a weakly augmented version of an unlabeled image, \(\mathrm{Aug}_w(x_j)\), and feed it through the model to obtain soft prediction logits \(q^{e}_{j,w}\). If the model is confident in this prediction determined by comparing its maximum probability to a predefined threshold $\tau = 0.95$, we convert these soft logits into a hard pseudo-label \(\hat q^{e}_{j}\).While recent approaches such as Adsh \cite{guo2022class} investigate adaptive thresholding strategies, we leave the exploration of such methods for future work and, in this paper, follow the original FixMatch \cite{sohn2020fixmatch} setting with a fixed threshold. We then create a strongly augmented version of the same image, \(\mathrm{Aug}_s(x_j)\), and enforce that the model’s prediction on this distorted image matches the pseudo-label \(\hat q^{e}_{j}\). This process forces the model to learn representations that are robust to significant augmentation, improving its consistency.

This unsupervised loss objective for an expert $e$ is expressed as the cross-entropy $\mathcal{H}$ between the pseudo-label \(\hat q^{e}_{j}\) and the prediction on the strongly augmented sample, averaged over all confident predictions:

\begin{equation}
\begin{aligned}
\mathcal{L}_u^e
&= \frac{1}{u} \sum_{j=1}^{u}
\mathbf{1}\bigl[\max q^{e}_{j,w}\ge\tau\bigr] \\
&\quad\times
\mathcal{H}\Bigl(
\hat q^{e}_{j},\,
\Phi_{\text{ex}}\bigl(
\Phi_{\text{emb}}(\mathrm{Aug}_s(x_j)),\,
\boldsymbol{\psi}_{e}
\bigr)
\Bigr)
\end{aligned}
\label{eq:Lu}
\end{equation}

\paragraph{Meta-objective.}
The overall loss aggregates across experts:
\begin{equation}
\mathcal{L}
  =\sum_{e=1}^{E}
      \Bigl(
        \mathcal{L}_{s}^{e}
        +\lambda\,\mathcal{L}_{u}^{e}
      \Bigr),
\label{eq:total-loss}
\end{equation}
where \(\lambda\) balances supervised and unsupervised terms.  
Optimising \(\mathcal{L}\) yields a shared encoder \(\Phi_{\text{emb}}\) and expert-specific embeddings \(\boldsymbol{\psi}_{e}\), that enables robust label prediction even for experts with limited initial labels.

\subsection{Context-Aware Expert Label Generation}\label{sec:inference}
Once the model is trained, we deploy it to achieve our primary objective: generating a complete set of context-aware expert labels for the entire dataset.  Given an expert context set \(\mathcal{C}_{e}\) and a query image \(x_{j}\), we first predict binary correctness of the expert on query image $x_j$:
\begin{equation}
\hat{h}_{\text{bin},j}^{e}
  =\arg\max
    \Phi_{\text{ex}}\!\bigl(
      \Phi_{\text{emb}}(x_{j}),\,
      \boldsymbol{\psi}_{e}
    \bigr).
\label{eq:binary-pred}
\end{equation}
Then the final categorical expert label is given by:
\begin{equation}
\hat{h}_{j}^{e} =
  \begin{cases}
    y_{j}, &\text{if }\hat{h}_{\text{bin},j}^{e}=1,\\[0.4em]
    \mathrm{Uniform}\bigl(\mathcal{Y}\setminus\{y_{j}\}\bigr), &\text{otherwise.}
  \end{cases}
\label{eq:label-gen}
\end{equation}
Applying this to all \((x_{j})\in D_{\text{gt}}\) for every expert \(e\) yields a dataset of augmented labels for the population $E$.

\subsection{Context-aware L2D}
\label{sec:l2d_pop}
The generated pseudo-labels provide the necessary supervision for our downstream L2D model. For this purpose, we adapt the L2D-Pop architecture \cite{tailor2024learning}, which personalizes deferral decisions by conditioning on each individual expert's context-set. 

Formally, based on the surrogate loss framework of \cite{husseinmozannar2020consistent}, the label set $\mathcal{Y}$ is augmented with a deferral option $\bot$, allowing the model to learn both classification logits $\mathbf{g}=(g_1,\dots,g_K)$ and a deferral logit $g_{\bot}$.
L2D-Pop achieves personalisation by encoding an expert's \emph{context set} $\mathcal{C}_e$, a small number of past decisions, into a permutation-invariant embedding $\boldsymbol{\psi}_e$. This embedding then conditions the deferral logit, $g_{\bot}(x,\boldsymbol{\psi}_e)$, enabling the model to tailor its deferral strategy to each specific expert. The population-aware surrogate loss is given as follows:
\begin{equation}
\begin{aligned}
\phi_{\text{L2D-Pop}}
&=
  -\log\frac{e^{g_{y}(x)}}{Z\!\bigl(x,\boldsymbol{\psi}_e\bigr)} \\[4pt]
&\quad\;\;
  - \mathbf{1}[m_{e}=y]\,
    \log\frac{e^{g_{\bot}\!\bigl(x,\boldsymbol{\psi}_e\bigr)}}{Z\!\bigl(x,\boldsymbol{\psi}_e\bigr)},
\end{aligned}
\label{eq:l2d-pop-loss-final}
\end{equation}
where
$Z\!\bigl(x,\psi_e\bigr)=e^{g_{\bot}(x,\psi_e)}+\sum_{k=1}^{K}e^{g_{k}(x)}$.
\medskip

We re-use the $\Phi_{\text{enc}}$ (Figure \ref{fig:schema}) to train the L2D-Pop model, as it represents the meta behaviour of the expert population.

\paragraph{Test‐time inference.}
At test time, for each expert \(e\) we first infer their context‐set embedding \(\psi_e\) by encoding the few labeled examples \(\mathcal{C}_e\) through the frozen encoder \(\Phi_{\mathrm{enc}}\).  Given a new query \(x^*\), the L2D-Pop head produces both classification logits \(\{g_k(x^*)\}_{k=1}^K\) and the deferral logit \(g_{\bot}(x^*,\psi_e)\).  The final decision is
\[
\hat{d} =
\begin{cases}
\displaystyle \arg\max_{k\in\{1,\dots,K\}} g_k(x^*),
  & 
    \substack{
      \text{if }g_{\bot}(x^*,\psi) \\ 
      < \max_k g_k(x^*)
    },
\\[1em]
\displaystyle \bot,
  &
    \text{otherwise}.
\end{cases}
\]
We thus defer to expert \(e\) whenever its conditional deferral score exceeds the highest class score, yielding personalized deferral decisions at inference time.

\section{Experiments}
\label{sec:experiments}

Our empirical study examines two questions: (1) Can a handful of past decisions per expert be leveraged to produce a large pool of high-quality synthetic labels, and (2) do those labels enable a downstream Learning-to-Defer model to generalise and defer effectively to unseen experts, achieving performance close to an oracle trained on all true expert labels? 

To this end, we first generate a complete set of synthetic expert labels from scarce initial data for the entire population. We then use these labels to train a downstream L2D-Pop model. We thereby demonstrate that our generated labels are of sufficiently high quality to improve this downstream task to near-oracle performance, even when adapting to experts unseen during training.

\subsection{Label Generation}

\paragraph{Datasets and Feature Backbone for Label Generation}
Our label generation framework is evaluated on three standard vision benchmarks: \textsc{CIFAR-10} (10 classes), \textsc{FashionMNIST} (10 classes), and \textsc{GTSRB} (43 classes). For this stage, a \textit{Wide-ResNet-28-10} network is used as the image encoder \(\Phi_{\text{emb}}\). For \textsc{CIFAR-10} and \textsc{FashionMNIST}, \(\Phi_{\text{emb}}\) is pre-trained on an 80\%/20\% train/validation split of the official training data. For \textsc{GTSRB}, it is trained on the full official training set and validated on the official test set. After pre-training, the backbone is frozen and used as a fixed feature extractor for the core task of modeling expert behaviour.

\begin{figure*}[t]
  \centering
  \includegraphics[width=\textwidth]{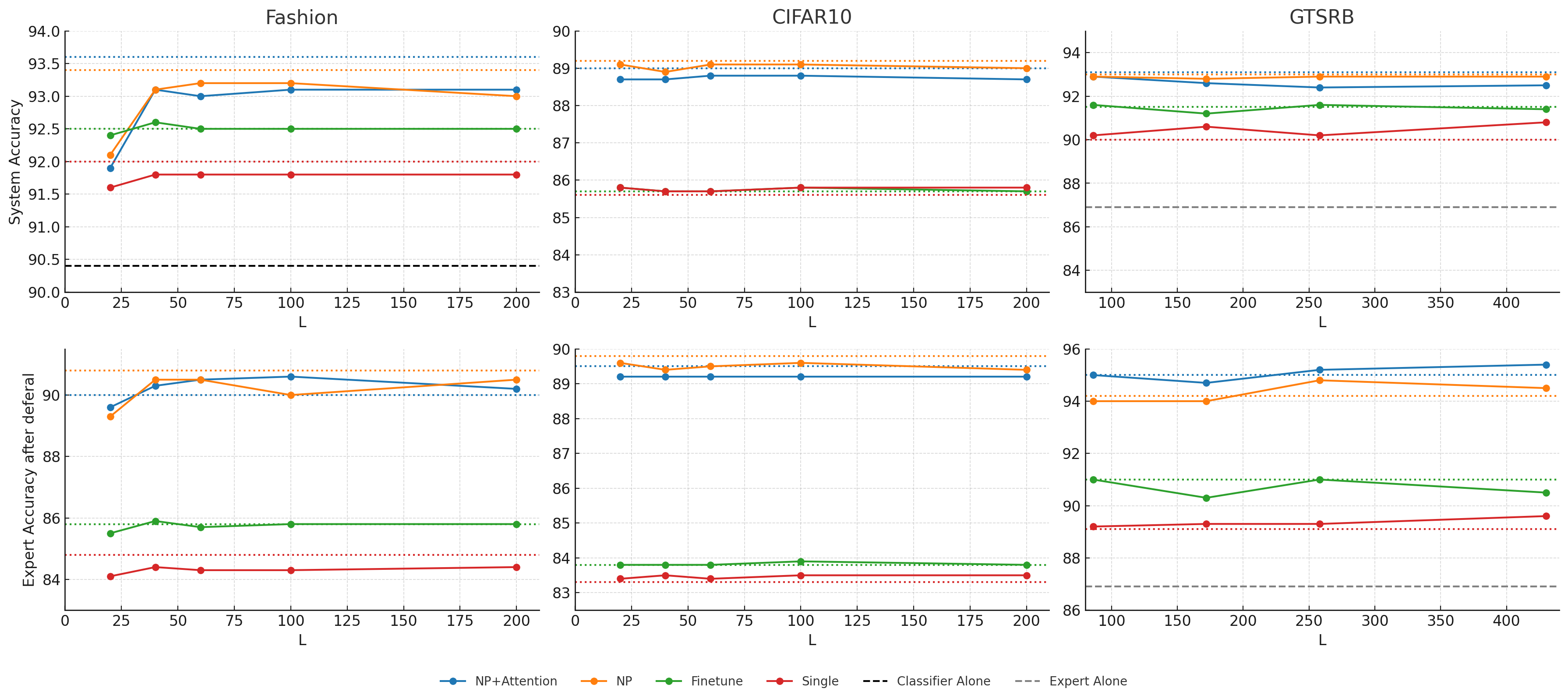}
  \caption{Impact of increasing \(L\) on downstream L2D-Pop performance.  
  Each column corresponds to one dataset—Fashion-MNIST (left), CIFAR-10 (center), and GTSRB (right).  
  \textbf{Top row:} overall human–AI system accuracy.  Solid lines show the four model variants (NP + Attention, NP, Finetune, and Single); matching-color dashed lines denote their oracle upper bounds, and the black dashed line marks the classifier-alone baseline.  
  \textbf{Bottom row:} expert accuracy after deferral.  Solid lines again show the four variants with their oracle upper bounds (matching-color dashed lines); the grey dashed line indicates the expert-alone baseline.}
  \label{fig:res}
\end{figure*}

\paragraph{Synthetic Expert Population}
To ensure a controlled and reproducible environment, we create a population of ten synthetic experts. Each expert is defined by an \textbf{oracle set} of classes it labels with 100\% accuracy. For our 10-class datasets, the oracle set size is \(H=8\), and for \textsc{GTSRB}, it is \(H=34\), establishing a baseline expert accuracy of approximately 80\%. For any input outside its oracle set, the expert provides a label chosen uniformly at random from the incorrect options. In practice this balance is critical: if experts are too weak the system rarely defers; if they are nearly perfect the classifier becomes redundant. With these H values we obtain a regime in which deferral is both meaningful and beneficial, as reflected in the accuracy gains reported across different expert strength for CIFAR10. To create a diverse population with distinct but overlapping skills, the oracle sets are generated \textbf{cyclically}, ensuring that the similarity between any two consecutive experts is controlled and providing a challenging testbed for personalization.

\paragraph{Number of Labels}
The label-generation model is trained to predict each of the ten experts’ correctness. We simulate per-expert data scarcity by limiting the number of available ground-truth annotations to \(L = N_{c}\times k\), where \(N_{c}\) is the number of classes and \(k \in \{2, 4, 6, 10, 20, 50, 250\}\) represents the \textbf{number of labeled examples per class}. In the most extreme case (\(k=2\) on \textsc{CIFAR-10}), the model must learn an expert's behaviour from only \(L = 20\) examples. During this training stage, the context-set size is fixed to \(B = 2 \times N_{c}\), matching the available data in the most data-scarce setting.

\begin{table*}[ht!]
  \centering
  \caption{Accuracy gain (\%) over the classifier baseline
           (76.3\%) on \textsc{CIFAR-10} for several expert strengths \(H\).
           The recommended setting \(H{=}8\) for CIFAR-10 is highlighted.}
  \label{tab:h_difference}
  \begin{tabular}{@{}c S[table-format=-2.1] S[table-format=-2.1] S[table-format=-2.1] S[table-format=-2.1]@{}}
    \toprule
    \textbf{Expert strength \(H\)} &
    \textbf{L2D-Pop(NP+Attention)} &
    \textbf{L2D-Pop(NP)} &
    \textbf{Finetune} &
    \textbf{Single-L2D} \\
    \midrule
      2 & -3.6 & -0.9 & -12.3 & -11.8 \\
      5 &  4.9 &  4.4 &  -4.1 &  -3.8 \\
    \textbf{8} & \textbf{12.7} & \textbf{12.9} & \textbf{9.4} & \textbf{9.3} \\
    \bottomrule
  \end{tabular}
\end{table*}

\subsection{Context-Aware L2D Training and Evaluation}
We choose the state-of-the-art L2D-Pop model proposed in ~\cite{tailor2024learning} as our base model.
A key aspect of this setup is the intentional use of a small, fallible CNN as the classifier’s feature extractor, following a common practice in L2D literature~\cite{tailor2024learning,husseinmozannar2020consistent,10.5555/3666122.3666281} This is mainly because datasets like \textsc{CIFAR-10} are largely solved, a high-capacity backbone would create a near-perfect classifier and render the deferral task trivial. This simpler model, consisting of only two convolutional blocks followed by a linear classifier head, ensures a meaningful testbed where both AI and expert have unique weaknesses. 
Using this backbone, we train two systems for comparison: the \textbf{Proposed System}, which uses our synthetic labels, and an \textbf{Oracle Upper Bound}, which uses the complete set of true expert labels. The models are trained using a 90\%/10\% train/validation split for \textsc{CIFAR-10} and \textsc{FashionMNIST} and the official test sets are used as testing set; for \textsc{GTSRB}, the official training set is used for training, and the official test set is split 50/50 for validation and testing. Finally, during evaluation, the system is evaluated on ten experts—\textbf{five ``seen'' experts} from the label-generation phase and five ``unseen'' experts whose behaviour must be inferred at test time—using a context-set size of $B=50$ as in the original L2D-Pop pipeline \cite{tailor2024learning}.

\subsubsection{Single-L2D and L2D–Pop variants}
We evaluate \textbf{Single-L2D}, a population-agnostic baseline trained on pooled expert data and three variants of the base L2D–Pop architecture: 
\textbf{NP+Attention}, our full model featuring a Neural Process encoder with multi-head attention; 
\textbf{NP}, a lighter variant in which the attention module is replaced by an MLP-based aggregator; 
\textbf{Finetune}, which initialises \textbf{Single-L2D} and fine-tunes it on each expert’s context set.

\subsubsection{Metrics}
We report two key metrics shown in Figure~\ref{fig:res}: \textbf{system accuracy}, which measures the performance of the complete system, and \textbf{Expert accuracy on deferred instances}, which is an indicator evaluating the quality of the deferral policy. 
By varying the initial data budget $L$, we quantify how effectively our synthetic labels close the performance gap to the oracle upper bound.

\section{Results}
\label{sec:results}
Figure~\ref{fig:res} summarises downstream system and expert accuracy as a function of the label budget $L$.
Across all three dataset, even a modest number of initial labels enables every L2D–Pop variant to surpass both stand-alone baselines and to approach its oracle upper bound.

\paragraph{System Performance Analysis}
For both the Fashion and CIFAR10 datasets, all tested methods demonstrate high system accuracy even with a limited budget of L=100 labels, achieving results close to their respective upper bounds. On the Fashion dataset, the L2D-POP variants NP+Attention and NP achieve system accuracies of 93.1\% and 93.2\%, which represents a significant increase of 2.7 and 2.8 percentage points (pp) over the standalone classifier's 90.4\% accuracy. The Finetune and Single methods also outperform the classifier, with accuracies of 92.5\% (a 2.1 pp improvement) and 91.8\% (a 1.4 pp improvement). This trend is even more pronounced on the CIFAR10 dataset, where NP+Attention (88.8\%) and NP (89.1\%) achieve substantial gains of 12.5 and 12.8 pp over the 76.3\% accuracy of the standalone classifier. The Finetune and Single methods also show strong performance, both reaching 85.8\% accuracy, a 9.5 pp improvement. Similarly, on the GTSRB dataset, with a low budget of L=86, all methods surpass the classifier-alone accuracy of 74.8\%. The NP+Attention and NP methods reach 92.9\% (an 18.1 pp gain), while Finetune achieves 91.6\% (a 16.8 pp gain) and Single reaches 90.2\% (a 15.4 pp gain).

\paragraph{Expert Deferral Performance}
In terms of leveraging expert input, all methods consistently improve upon the expert's standalone accuracy. For the Fashion dataset at L=100, the NP+Attention and NP methods achieve expert accuracies of 90.6\% and 90.0\%, marking an 8.6 and 8.0 pp increase, respectively, over the expert's baseline of 82.0\%. Finetune and Single also show improvements of 3.8 and 2.3 pp, respectively. On the CIFAR10 dataset, the L2D-POP variants continue to lead, with NP+Attention (89.2\%) and NP (89.6\%) improving upon the expert by 7.2 and 7.6 pp. Finetune and Single methods provide more modest gains of 1.9 and 1.5 pp. For the GTSRB dataset, with a budget of L=86, the NP+Attention method achieves an expert accuracy of 95.0\%, which matches its upper bound and is 8.1 pp higher than the expert alone (86.9\%). The NP, Finetune, and Single methods also demonstrate effective deferral, improving expert accuracy by 7.1, 4.1, and 2.3 pp, respectively.

\section{Discussion}
\paragraph{High-Quality Labels from Scarce Data.}
The most salient result is the data efficiency of our framework. Across all three datasets, the performance of the system trained on synthetic labels (solid orange and red curves) rapidly approaches the oracle upper bound (dashed black and purple lines) with only a small number of initial expert annotations, \(L\). With as few as \(L=50\) labels per expert (e.g., \(k=5\) for 10-class datasets), our system already closes most of the performance gap. Furthermore, the overall system accuracy consistently surpasses the performance of either the classifier-alone or expert-alone baselines, demonstrating that our generated labels successfully enable an effective AI–expert collaboration. Notably, these strong results hold for both experts seen during label generation and for completely novel experts, demonstrating the generalization capability of our approach. 
\paragraph{Sensitivity to Initial Label Quality.}
A key limitation of our approach is its sensitivity to label noise in the small initial dataset ($L$). Our semi-supervised method assumes consistent labeling for similar inputs, a principle that inconsistent expert errors can violate. Consequently, the framework requires a "clean" initial set—where an expert's behavior for any given input type is consistently correct or incorrect—as any noise in this small seed set can disproportionately degrade performance. This requirement presents a clear trade-off. Unlike methods such as PL2D \cite{nguyen2025probabilistic} that handle noisy annotations by using more data, our approach is optimized for data-scarce scenarios. It achieves strong performance with very few labels (e.g., $L=50$), provided this initial set is of high quality. Our method thus exchanges robustness to label noise for higher data efficiency.

\paragraph{Sensitivity to expert strength $\mathcal{H}$}Table~\ref{tab:h_difference} shows how system accuracy on \textsc{CIFAR-10} changes, relative to a 76.3\% classifier baseline, as the simulated expert’s strength \(H\) increases. With a weak expert (\(H=2\)), deferring offers little benefit—\textbf{L2D-Pop(NP+Attention)} still falls 3.6\,pp below the baseline, and population-agnostic models fare even worse. Once the expert reaches moderate reliability (\(H=5\)), the two \textbf{L2D-Pop} variants turn this deficit into gains of about +4.5\,pp, whereas \textbf{Finetune} and \textbf{Single-L2D} remain negative. When the expert is strong (\(\mathbf{H=8}\)), every method improves markedly; the expert-conditional \textbf{L2D-Pop} variants lead with gains of +12.7\,pp and +12.9\,pp, clearly outperforming simple fine-tuning and population-agnostic training. 

\section{Conclusion}
This paper addressed the expensive data requirements that hinder the practical application of population-aware learning to defer systems. Our work introduces a novel, context-aware semi-supervised framework that uses meta-learning to generate high-quality synthetic labels for a diverse expert population from only a few initial demonstrations. Extensive experiments show that a downstream model trained on our pseudo-labels approaches oracle-level performance and generalizes effectively to unseen experts. While reliant on a small, high-quality set of limited annotations, our method significantly lowers the barrier for deploying sophisticated, adaptive L2D systems in the real world. By resolving this critical training data bottleneck, we pave the way for more practical, scalable, and robust human-AI collaboration in critical decision-making environments.

\section*{Acknowledgment}

G.C. was supported by the Engineering and Physical Sciences Research Council (EPSRC) through grant EP/Y018036/1.

\bibliographystyle{IEEEtran}   
\bibliography{ref}       

\end{document}